\documentclass[pre,numerical,a4paper,superscriptaddress,showpacs]{revtex4-1}
\usepackage{mycommands}
\begin{document}
\title{Modeling the Mechanosensitivity of Fast-Crawling Cells on Cyclically
  Stretched Substrates}
\author{John J. Molina} \email{john@cheme.kyoto-u.ac.jp}
\affiliation{Department of Chemical Engineering, Kyoto University, Kyoto 615-8510}
\author{Ryoichi Yamamoto}
\affiliation{Department of Chemical Engineering, Kyoto University, Kyoto 615-8510}
\affiliation{Institute of Industrial Science, The University of Tokyo,
Tokyo 153-8505}
\date{\today}
\begin{abstract}
The mechanosensitivity of cells, which determines how they are able to
respond to mechanical signals received from their environment, is
crucial for the functioning of all biological systems. In experiments,
cells placed on cyclically stretched substrates have been shown to reorient in a
direction that depends not only on the type of cell, but also on the
mechanical properties of the substrate, and the amplitude and rate of
stretching. However, the underlying biochemical and mechanical
mechanisms responsible for this realignment are still not completely
understood. In this study, we introduce a computational model for
fast crawling on cyclically stretched substrates that accounts for the
sub-cellular processes responsible for the cell shape and motility, as
well as the coupling to the substrate through the focal adhesion
sites. In particular, we focus on the role of the focal adhesion
dynamics, and show that the reorientation under cyclic stretching is
strongly dependent on the frequency, as has been observed
experimentally. Furthermore, we show that an asymmetry during the
loading and unloading phases of the stretching, whether coming from
the response of the cell itself, or from the stretching protocol, can
be used to selectively align the cells in either the parallel or
perpendicular directions.
\end{abstract}
\maketitle
\section{Introduction}
\label{s:intro}
The structure and function of cells is carefully regulated by the signals
they receive from their environment. Of particular interest is the
transfer of mechanical forces and stresses, which in turn are known to
trigger specific bio-chemical responses inside the cell that can
significantly alter their behavior, inducing changes in shape, size,
motility, reorganization of the cytoskeleton, and even cell
proliferation and differentiation\cite{Janmey2007,Crowder2016}. This last example is probably the
most striking, given the bio-medical applications it
promises. Carefully engineered bio-materials should allow us to control
stem cell fate decisions, i.e., whether or not they divide or
differentiate, and which specific cell lineage is
chosen\cite{Hiew2018}. However, before this is possible, we need to
have a fundamental understanding of the interactions between the cells and
the chosen bio-material.

One of the preferred methods to probe the mechanical interaction of
cells with their environment is to place them on an elastic
substrate that is being periodically stretched along a given
direction. Studying how the cells respond to this 
perturbation provides crucial information on its mechanosensing
abilities. Following Iwadate et
al.\cite{Iwadate2009,Iwadate2013,Okimura2016, Okimura2016a}, it is
useful to distinguish between slow crawling cells, such as
fibroblasts, endothelial, and smooth muscle cells, and fast crawling cells such
as \textit{Dictyostelium} or neutrophil-like HL-60; where the typical
migration velocities can differ by one to two orders of magnitude
between the two types. For example, the average speed of
fibroblasts is of the order of $10\mu \text{m}/\text{h}$\cite{Ebata2018},
whereas \textit{Dictyostelium} can move at speeds on the order of
$10\mu \text{m}/\text{s}$\cite{Iwadate2009}. In addition, slow
crawling cells typically possess stress fibers, whereas fast
crawling cells do not. This is a crucial difference to understand
their mechanosensitive response.

Early experiments on fibroblasts\cite{Buck1980} and endothelial
cells\cite{Dartsch1989,Iba1991} found that these cells preferred to align
their stress fibers in a direction perpendicular to the
stretching. This
reorientation of the stress fibers has been linked to the
depolymerization and disassembly of parallel
fibers\cite{Hayakawa2001,Jungbauer2008}. Nevertheless, it is also possible
for the stress fibers to align parallel to the direction of stretching, as demonstrated
experimentally on endothelial cells with inhibited
Rho-kinase activity (which would tend to lower the myosin activity and
thus the base tension)\cite{Lee2010}. Finally, while the alignment
of the stress fibers is correlated with the cell reorientation, it is
by no means sufficient. This was shown by experiments on vascular smooth muscle cells, in which stress activated cation
channels where inhibited, resulting in cells that were randomly
oriented, even though they contained oriented stress
fibers\cite{Hayakawa2001}. 

The fact that cells without stress fibers also exhibit characteristic reorientation under
cyclic stretching is clear evidence that stress fiber realignment
cannot be the only mechanism responsible for the
reorientation. Unfortunately, the fast crawling nature of these cells
makes experimental observations much more difficult, since it requires that the cell motion be tracked. Indeed
it was only recently that the group of Iwadate managed to perform such experiments\cite{Iwadate2009,Iwadate2013,Okimura2016a,Okimura2016}.
They have found that \text{Dictyostelium} cells prefer to
migrate in the perpendicular direction. This occurs without any
ordering of the dense actin-network inside the cell, but it is
accompanied by the formation of dense myosin bundles at the lateral
edges, preventing any pseudopod extension in those directions. Further
experiments on other fast-crawling cells, such as HL-60 and Blebbistatin-treated
(stress fiber less) keratocytes also found similar perpendicular
alignment\cite{Okimura2016a,Okimura2016}. This response has yet to be
fully explained, and it is even less understood than the reorientation of slow-crawling cells, where the stress fibers seem to play a dominant role.

The mechanism responsible for the realignment is evidently cell-specific,
and likely to depend on the experimental conditions. However, it is 
clear that it should involve several general ingredients, namely, the
focal adhesion dynamics, through which the cell is able to transfer
forces to and from the substrate, and the actin network and myosin induced contractility, responsible for the migration
of the cell, as well as the mechanical properties of the
cytoskeleton. Alternative theories have been proposed that can explain
the reorientation as arising from one of these elements. For example,
as a consequence of the passively stored elastic
energy\cite{Livne2014a} or the forces on the focal
adhesions\cite{Zhong2011,Chen2015a}, with both theories capable of reproducing
the same experimental data, even though they are modeling different
mechanisms, under different assumptions. Furthermore, all such models
seem to have been developed with slow-crawling
cells in mind, where stress fibers are likely to play a crucial role,
and where the motion of the cells can be decoupled from their
reorientation. When considering fast crawling cells such as
\textit{Dictyostelium} or HL-60, the
motility of the cell can no longer be decoupled from its
reorientation. Thus, we must consider the dynamic remodeling of
the relevant sub-cellular elements (e.g., the actin-cytoskeleton and
the focal adhesions) under the cyclic stretching and how this affects
the motion of the cell. In such cases, theoretical approaches quickly
become intractable, and we must resort to computational modeling. 

In this work, we extend an established phase-field model of
crawling cells\cite{Lober2014} to describe the dynamics of
fast-crawling cells over substrates undergoing large amplitude cyclic
deformations. We then use this model to study the reorientation
dynamics as a function of frequency. Based on recent studies, which
report a strong frequency dependence for the stability of focal
adhesions\cite{Kong2008,Zhong2011}, we assume that the coupling to the
substrate is given by a frequency-dependent detachment rate. At low
frequencies, the cell and its constituent elements are able to follow
the deformation and no reorientation is observed. At moderate
frequencies, but below the threshold value that triggers the instability of
the focal adhesions, both parallel and perpendicular orientations are
stable. Increasing the frequency over this threshold, a
moderate frequency range is found over which only the perpendicular
direction is stable (as seen experimentally). Furthermore, by tuning the response of the
cells to detach only during fast extension, this realignment
effect can be strengthened, and it is entirely reversed to favor
parallel alignment if the detachment occurs during fast compression. We thus find that an
asymmetry in the cellular response during loading and unloading
can have a dramatic effect on their reorientation dynamics. This could
be tested experimentally by employing a non-symmetric stretching
protocol (i.e., fast extension accompanied by slow compression, and
vice versa). Finally, upon a
further increase in the frequency, only the parallel orientation is
stable. The observed realignment response depends on whether the frequency
of stretching is probing the shape deformation, the actin-network, or
the focal adhesion dynamics. While we have used a generic model for
crawling cells, and only including the response of the focal adhesion
sites to the stretching, the framework we propose can be easily
used with more elaborate phase-field models developed and parameterized
for specific cell types. This will allow us to investigate the
biomolecular and mechanical origins of the cell's mechanosensitive response
in much more detail.

\section{Model}
\label{s:model}
Ever since the pioneering studies of Cahn, Hilliard, and Allen, who
introduced phase-field models to study the phase-separation of
binary alloys\cite{Cahn1958,Cahn1961,Allen1972,Allen1973}, phase-field modeling has become one of the preferred
methods for physicists and material scientists to describe
microstructural dynamics in systems with non-homogeneous
``phases''. These phases can be used to represent any
material property of interest, from a difference in density, orientational order,
or chemical composition, to differences in electric or magnetic
polarization, thus providing a universal framework with which to study
a wide variety of phenomena. Recently, this approach has seen
considerable success outside of physics, and is now actively used to
address problems in biology and even medicine. Notable examples include,
among others, studies on the morphodynamics of crawling
cells\cite{Shao2010,Shao2012,Ziebert2012,Lober2014,Palmieri2015}, the immune response to invading
pathogens\cite{Najem2014}, axonal extension of nerve
cells\cite{Najem2013,Takaki2015}, and cartilage
regeneration\cite{Yun2013}, as well as tumor
growth\cite{Sciume2013,Lima2014}. In this work, we focus on the
mechanosensitivity of crawling cells, and in particular on their ability
to sense and respond to mechanical cues from a substrate undergoing cyclic
stretching. A phase-field approach is ideally suited for this purpose,
particularly for fast crawling cells, as it provides a cell-level description which can take into account
the acto-myosin based propulsion mechanism, the force transmission to
and from the substrate (mediated by the focal adhesion sites), as well
as allowing for the large shape deformations caused by the externally applied
strain. In addition, this type of modeling can easily scale upward to
consider the collective dynamics of multi-cellular systems and
confluent tissues\cite{Lober2015}. In this section, we will introduce the basic
phase-field model of crawling cells that we have adopted, which was
originally designed to describe the motion of keratocyte-like
fragments over viscoelastic substrates without any global deformation. Then, we
define the periodic strain imposed on the substrate, and
the extensions to the model that are required to consider crawling
under such large-amplitude cyclic deformations.
\subsection{Phase-Field Model of Cells on Viscoelastic Substrates}
We adopt the 2D model originally developed by Ziebert and Aranson\cite{Ziebert2012}, which
describes each cell using a non-conserved order parameter $\rho$,
whose values lie between zero (outside the cell) and one (inside the
cell). This allows for an implicit tracking of the boundary, avoiding
many of the computational difficulties of related sharp interface
methods. A free energy functional $F[\rho]$ is then associated to this
order parameter and determines the driving force for its time
evolution as
\begin{align}
  \p_t\rho &= -\Gamma \frac{\delta F[\rho]}{\delta \rho}\label{e:modela}
\end{align}
with $\Gamma$ the mobility coefficient for $\rho$. This is the
so-called ``Model A'' or time-dependent Ginzburg-Landau model\cite{Chaikin1995}. To
lowest order, the free energy functional takes the form
\begin{align}
  F[\rho] &= \int\barg[\big]{f(\rho) + D_\rho\parg{\Grad\rho}^2} \vdf{x}\label{e:Frho}
\end{align}
where $f(\rho)$ is the free energy density of the  homogeneous system
and the term proportional to $D_\rho$ provides a penalty term to the
formation of sharp interfaces. The free-energy density is defined to have a
double-well form, representing the local stability of the two phases
$\rho=0$ and $\rho=1$, and is given by
\begin{align}
  f(\rho) &= \int_0^\rho\parg{1-\rho^\prime}\parg{\delta[\rho] - \rho^\prime}\rho^\prime\df{\rho^\prime}\label{e:frho}
\end{align}
where the value of $\delta$ controls the relative stability of the two. The motility of the cell is modeled by introducing
an additional polar order parameter $\bm{p}$, which gives the average
orientational order of the actin filament network responsible for the
motion. These filaments are continuously polymerizing at the leading
edge and pushing against the membrane, allowing the cell to extend
forward. This requires that the cell be able to transfer the
forces to the substrate, something it is able to due because the
actin-network is connected to the substrate through the focal adhesion
bonds. This is modeled by introducing an additional scalar field $A$,
representing the density of adhesion bonds. Finally, the coupled set of equations for $\rho$, $\bm{p}$, and
$A$ are given by\cite{Lober2014}
\begin{align}
  \partial_t \rho &= D_\rho \nabla^2 \rho - \parg[\big]{1-\rho}\parg[\big]{\delta[\rho] - \rho}\rho - \alpha(A)\bm{\nabla}\rho\cdot \bm{p} \label{e:rho0}\\
\partial_t \bm{p} &= D_{p}\nabla^2\bm{p} - \tau_1^{-1}\bm{p} - \tau_2^{-1}\parg[\big]{1-\rho^2}\bm{p} - \beta f\barg[\big]{\bm{\nabla}\rho} - \gamma(\bm{\nabla}\rho\cdot\bm{p})\bm{p} \label{e:p0}\\
\partial_t A &= D_A \nabla^2 A + \rho\parg[\big]{a_0\Norm{\bm{p}} + a_{\text{nl}} A^2} - \parg[\big]{d(u) + sA^2} A\label{e:A0}
\end{align}
where, without loss of generality we have taken $\Gamma = 1$. For the
dynamics of $\rho$ (Eq.\eqref{e:rho0}), the first two terms on the
right-hand side result from taking the functional
derivative of the energy functional of Eq.~\eqref{e:Frho}, while the
last term, proportional to $\Grad\rho\cdot\bm{p}$, and akin to an
advection term, represents the active contribution of the
actin-network pushing the cell membrane. The strength with which the
actin network can push on the membrane is given as a function of the
local density of adhesion sites $\alpha(A) = \alpha \cdot A$. The dynamics of $\bm{p}$ (Eq.\eqref{e:p0})
is given by a simple reaction-diffusion equation, with a source
term to account for the polymerization at the interface ($\propto
\beta f[\Grad\rho]$), and a decay term ($\propto \tau_1^{-1}$) to account
for the corresponding depolymerization. The polymerization rate
is chosen to be a function of the gradient of $\rho$ that ensures that
the growth rate is bounded and limited to the interface, with
\begin{align}
  \bm{f}[\bm{x}] &= \frac{\bm{x}}{\sqrt{1 + \epsilon\Norm{\bm{x}}^2}}\label{e:fx}
\end{align}
As such, the maximum growth rate is given by $\simeq \beta/\sqrt{\epsilon}$.
Note that an additional decay
term $\propto \tau_2^{-1}(1-\rho^2)$ is included for computational
simplicity, to make sure that the actin field is non-zero only inside
the cell. The last term in Eq.~\eqref{e:p0} accounts for the myosin
induced bundling at the rear of the cells\cite{Ziebert2012}, helping to break the
$\pm\bm{p}$ symmetry and favor polarization. A similar reaction-diffusion model is used for the
concentration of adhesion sites $A$~(Eq.~\eqref{e:A0}). Naturally, the
attachment to the substrate can only occur inside of the cell: there
is a linear term  proportional to $a_0 \Norm{p}^2$, since the attachments
require the presence of actin, and a non-linear term
$a_{\text{nl}} A^2$ to model the maturation and growth of existing
bonds. For the detachment, there is a linear term that couples the
dynamics of $A$ with the substrate displacement $u$, and a non-linear
term that saturates the total number of bonds. Finally, the $\delta$
function controlling the relative stability of the two phases is
given by ($\avg{\cdot} = \int\cdot\,\vdf{r}$)
\begin{align}
  \delta[\rho] &= \frac{1}{2} + \mu\parg[\big]{\avg{\rho} - \pi r_0^2}  - \sigma\Norm{\bm{p}}^2\label{e:delta0}
\end{align}
where the second term on the right hand side acts as a global
constraint on the cell volume (with $r_0$ the radius of the
non-polarized static cell), and the third term accounts for the
myosin-induced contraction.

At first glance, the model can seem overwhelming, as it possesses
over a dozen free parameters. Fortunately, a detailed analysis of this
model and its variants has already been
performed\cite{Aranson2016,Ziebert2016}, allowing us to focus on the
few parameters relevant for a study on the mechanosensitivity of cells on
cyclically stretched substrates. The activity of the cell can be
controlled by the strength of the propulsion ($\alpha$) and the rate
of polymerization ($\beta$). The shape of the cell can be controlled
mainly by the strength of the contractility ($\sigma$), with low (high)
values resulting in fan(crescent)-like shapes. The motor-asymmetry
($\gamma$) has only a small effect on the shape or dynamics of the
cell and can be considered constant without loss of
generality. Of the remaining parameters appearing in the equations of motion for
$\bm{p}$ and $A$, the most important is $a_0$, which sets the rate at
which new adhesion sites can be formed with the substrate. For
example, to consider patterned substrates, one would make this
parameter be position dependent. Such a study has been presented in
Ref.~\cite{Lober2014}, where a viscoelastic Kelvin-Voigt model is
used to describe the displacement of the substrate due to the traction
forces exerted by the cell. By controlling just two parameters, the
stiffness of the substrate and the rate of attachment ($a_0$), the
authors report a wide variety of motility modes, such as steady
gliding motion, stick-slip, bipedal and wandering, which have also
been observed experimentally\cite{Barnhart2010,Riaz2016}.

\subsection{Substrate Deformation}
We consider a substrate that is being cyclically stretched along one
of its axes. In most cases, this will necessarily imply a
compression along the perpendicular axes, with an amplitude that
depends on the Poisson's ratio $\nu$ of the material. To describe this deformation, it is convenient to introduce
Lagrangian (material) coordinates $\bm{\xi}$ to label the substrate
elements. The time-dependent (Eulerian) coordinates of a given element
$\bm{\xi}$ are then given by $\bm{x} = \bm{x}(\bm{\xi},t)$, which, for
the present case is given explicitly by
\begin{align}
  x^1 &= \parg[\big]{\xi^1 - L_x/2} \parg[\big]{1 + \eps(t)} \label{e:x1}\\
  x^2 &= \parg[\big]{\xi^2 - L_y/2} \parg[\big]{1 + \eps(t)}^{-\nu} \label{e:x2}
\end{align}
where $\eps$ is the lateral strain (along which the substrate is being
actively deformed), and $L_x$ and $L_y$ are the (undeformed) substrate
dimensions. For simplicity, we assume a sinusoidal perturbation given by
\begin{align}
  \eps(t) &= \frac{\eps_0}{2}\parg[\big]{1 - \cos{\parg{2\pi \omega t}}} \label{e:eps}
\end{align}
We thus have two equivalent representations for our system, in terms of
the body ($\bm{\xi}$) or lab ($\bm{x}$) frame. Given the time-dependent
deformation of the substrate, it is more convenient to solve the
equations of motion in the body frame, which is by definition
constant, than it is to solve them in the lab
frame. This is a common strategy when solving flow or elasticity
problems in the presence of time-dependent boundary conditions\cite{Luo2004,Venturi2009,Molina2016}. However, this
requires careful consideration, particularly with regards to the definition of the time derivatives.

Let $\bm{e}_i$ and $\bm{E}_I$ be the basis vectors in the lab and body frame,
respectively, and $u^i$ and $u^I$ the corresponding (contravariant) components of a
given vector $\bm{u} = u^i \bm{e}_i = u^I \bm{E}_i$. Throughout this work we will
assume the Einstein summation convention, and reserve lower (upper) case
indices for quantities in the lab (body) frame. The corresponding
transformation rules are given by\cite{Schutz1980}
\begin{align}
  \bm{e}_i &= \Lambda^{I}_{\phantom{I}i} \bm{E}_I  & \bm{E}_I &=
  \Lambda^{i}_{\phantom{i}I}\bm{e}_i \label{e:Fei}\\
  u^i &= \Lambda^{i}_{\phantom{i} I} u^I &
  u^I &= \Lambda^{I}_{\phantom{I} i} u^i\label{e:Fvi}
\end{align}
with $\Lambda^I_{\phantom{I}i} \equiv \p\xi^I/\p x^i$,
$\Lambda^i_{\phantom{i}I} \equiv \p x^i/\p\xi^I$, and
$\Lambda^{I}_{\phantom{I}i}\Lambda^{i}_{\phantom{i}J} =
\delta^{I}_{\phantom{I}J}$. The
inner or scalar product between two vectors is defined as $\vec{u}\cdot\vec{v}
\equiv u_I v^J = u^I v_J = G_{IJ} u^I v^J = G^{IJ} u_I v_J$, with
$G_{IJ}$ and $G^{IJ}$ the components of the metric tensor and its
inverse ($G^{IJ} G_{JK} = \delta^{I}_{\phantom{I}K}$)
\begin{align}
  G_{IJ} &= \Lambda^{i}_{\phantom{i}I}\Lambda^{j}_{\phantom{i}J}
  g_{ij} = \begin{pmatrix}
    \parg{1 + \eps(t)}^2 & 0 \\
    0 & \parg{1 + \eps(t)}^{-2\nu}
  \end{pmatrix}\label{e:GIJ}
\end{align}
where the metric tensor in the lab frame is the Euclidean metric
tensor $g_{ij} = \delta_{ij}$. For what follows, we will also require
the coordinate flow velocity $\bm{U}$, i.e., the velocity of the
coordinates or the velocity of the moving substrate. In the body frame, this is defined as\cite{Venturi2009}
\begin{align}
  \bm{U} &\equiv -\frac{\p\bm{\xi}}{\p t} =  \begin{pmatrix}
    \phantom{-\nu}\epst \parg[\big]{\xi^1 - L_x/2} \\
    -\nu \epst \parg[\big]{\xi^2 - L_y/2}
  \end{pmatrix} \label{e:U}
\end{align}
where $\epst = \frac{\dot{\eps}}{1 + \eps}$ and
\begin{align}
  \dot{\eps}(t) &=\p_t \eps(t) = 2\pi\omega\frac{\eps_0}{2}\sin{\parg{2\pi \omega t}}\label{e:doteps}
\end{align}

\subsection{Crawling Cells on Cyclically Stretched Substrates}
To consider the dynamics of the cell on the cyclically stretched
substrate, we begin by writing down the equations of motion in
contravariant form in the body (substrate) frame of reference,
replacing the time-derivatives with intrinsic time derivatives (see
Appendix~\ref{s:app_tensor}), to obtain
\begin{align}
  \p_t\rho  &= D_\rho \Delta \rho -
  \parg[\big]{1-\rho}\parg[\big]{\delta[\rho]-
    \rho}\rho - \alpha\parg{A} p^J \covd{\rho}{J} \label{e:rho}\\
  \p_t p^I &= D_p \Delta p^I - \tau_1^{-1}p^I -
  \tau_2^{-1}\parg{1-\rho^2}p^I - \beta G^{IJ} \frac{\covd{\rho}{J}}{1 +
    \epsilon \cntd{\rho}{K} \covd{\rho}{K}} - \gamma \parg[\big]{p^J
    \covd{\rho}{J}} p^I - p^J \grad_J U^I\label{e:p}\\
  \p_t A &= D_A\Delta A -\tau_A^{-1}(1-\rho^2)A + \rho\parg[\big]{a_0 p^J p_J +
    a_{\text{nl}}A^2} - \parg[\big]{d(\cdots) + s A^2}A - A\grad_J
  U^J \label{e:A}
\end{align}
where $\Grad_J \rho = \p_{\xi^J} \rho = \p_I \rho$ and $\grad_J U^I = \p_{J}
U^I + \Gamma^{I}_{KJ} U^K$ are the components of the covariant derivative
of $\rho$ and $\bm{U}$, respectively ($\Gamma^{I}_{JK}$ the connection
coefficients). In addition, the Laplacian
operator $\grad^2$ is here replaced with the Laplace-Beltrami
operator $\Delta$. In the current case, all connection
coefficients are zero ($\Gamma^{I}_{JK} = 0$), considerably
simplifying the calculations, since $\Delta \rho =
G^{JK}\p_{\xi^J}\p_{\xi^K}\rho$ and $\Delta p^I =
G^{JK}\p_{\xi^J}\p_{\xi^K} p^I$. The final set of equations
(\ref{e:rho}-\ref{e:A}) are almost the same as in the original
formulation (\ref{e:rho0}-\ref{e:A0}), except for the last term on the
right-hand side of the equations for $\bm{p}$ and $A$, which depends on
the gradient of the coordinate flow velocity ($\Grad\bm{U}$)
\begin{align}
  \Grad\bm{U} &\equiv \begin{pmatrix}
  \grad_1 U^1 & \grad_2 U^1 \\
  \grad_1 U^2 & \grad_2 U^2
  \end{pmatrix} =
  \begin{pmatrix}
    \epst(t) & 0 \\ 0 & -\nu\epst(t)
  \end{pmatrix}\label{e:DU}
\end{align}
and an additional decay term ($\tau_A^{-1}$) for the adhesion sites outside the cell. We found the latter to be necessary to avoid
any spurious adhesion-mediated interactions between the cell and its
periodic images, particularly when using small system sizes or
low frequencies.
The precise functional form for the detachment rate $d$ will be
discussed in the next subsection.
The additional term in the equation for $\bm{p}$ comes from the
time-dependent nature of the basis vectors, whereas the term appearing
in the equation for $A$ comes from the time-dependence of the volume
element, and is required to ensure the total conservation of bonds under stretching.
Of note is the fact that the equations of motion are translationally invariant, i.e., there is no explicit dependence on
the coordinates $\bm{\xi}$. This allows us to assume periodic boundary
conditions and employ efficient pseudo-spectral methods to solve the
equations. Details on the numerical implementation can be found in Appendix~\ref{s:app_num}.
\subsection{Cell-Substrate Coupling}
In this work, we are interested in studying the 
reorientation of fast-crawling cells such as \textit{Dictyostelium},
which possess no stress fibers, on cyclically stretched
substrates. Recent experiments by Iwadate et al.\cite{Okimura2016a} have shown that cell reorientation
occurs even though no significant orientational order is observed in
the dense actin-network in the middle of the
cell. Instead, the authors have reported that
myosin II becomes concentrated on the stretched sides of
the cell, but how this is related to the reorientation response, or
which pathway the cell uses to sense the mechanical stimulation, is
still not understood. However, they conclude their work by offering
three possibilities for how the mechanical signals trigger the
localization, (1) through the focal adhesion sites, (2) through some
unidentified mechanosensitive channel, or (3) through the deformation
of the actin filament network, among which they identify the latter as
more likely. Here, we will consider the first option, given the
obvious importance of the focal adhesions in the transmission of
forces to and from the cell, and the actin-network in
particular. Indeed, recent studies on
slow-crawling, stress fiber containing cells, have shown that the
adhesion dynamics can help to explain the experimentally observed
reorientation of such cells\cite{Zhong2011,Chen2015a}. 
Thus, for simplicity, we will ignore any effects coming
from the viscoelastic properties of the actin-network, even though
it surely has a role to play in determining the reorientation
response, particularly at lower
frequencies\cite{Kong2008,Zhong2011}. We therefore consider that the
coupling between the cell dynamics and the substrate is due
exclusively to the adhesion dynamics.

Under cyclic stretching, adhesion bonds have been shown to lose
stability if the frequency is high
enough\cite{Kong2008,Zhong2011}. This is due to the high speed
changes in the substrate, which prevent the formation of any stable
bonds. This frequency dependence for the stability of the adhesion
bonds has been linked to the strong frequency dependence of the
reorientation response seen experimentally. In particular, Liu et al.\cite{Liu2008} found that the alignment of arterial smooth muscle
cells is maximized for a given value of the stretching frequency, and
Jungbauer et al.\cite{Jungbauer2008} and Greiner et
al.\cite{Greiner2013} both reported a lower threshold
frequency below which no alignment is observed. Although it should be
noted that the former found the response time to decrease with increasing frequency
(above the lower threshold), before plateauing at an upper frequency
threshold, whereas the latter found no such change.

Within the phenomenological framework we are considering, we
incorporate this frequency dependent response in the form of a strain dependent
detachment rate. Based on the experimental results showing a
lower frequency threshold needed to observe any
realignment\cite{Jungbauer2008,Greiner2013}, and the strong frequency
dependence found for the stability of focal adhesions\cite{Kong2008},
we assume that the detachment rate is sensitive only to the rate at
which the substrate is being stretched. As an objective measure for
this rate of stretching, we use the (Lagrangian) rate of deformation tensor $\tensor{D}$, defined
as the time-derivative of the Green deformation tensor (or the right
Cauchy-Green tensor), which in
component form is given by\cite{Marsden1994}
\begin{align}
2 D^{I}_{\phantom{I}J} &= G^{K I} g_{i k}\parg[\bigg]{ \Lambda^{i}_{J}
  \grad_K U^k + \Lambda^{k}_{\phantom{k}K}\grad_J U^i }\label{e:Dij}
\end{align}
In Eulerian terms, it yields the symmetric part of the velocity
gradient tensor, and, as its name suggests, it provides
information on the rate at which an object is being deformed or
stretched. We consider that the rate of detachment $d$ depends solely on the
trace of this rate of deformation tensor $D=\trace{(\tensor{D})}$, i.e., how fast it is being
stretched or compressed. We assume a sharp sigmoidal response,
such that $d=0$ ($d=1$) below (above) the critical frequency
$\omega_c$. We introduce three basic response functions
\begin{align}
  d^{(\pm)}(D) &= \frac{d_0}{2}\barg[\bigg]{1 + \tanh{\parg[\Big]{{b^2\parg[\big]{D^2
            - D_c^2}}}}} \label{e:Dpm}\\
  d^{(+)}(D) &= \frac{d_0}{2}\barg[\bigg]{1 + \tanh{\parg[\Big]{{b^2\parg[\big]{R^2(D)
  - D_c^2}}}}}\label{e:Dp}\\
  d^{(-)}(D) &= \frac{d_0}{2}\barg[\bigg]{1 + \tanh{\parg[\Big]{{b^2\parg[\big]{R^2(-D)
  - D_c^2}}}}}\label{e:Dm}
\end{align}
with $d_0$ the maximum rate of detachment, $D_c$ the critical deformation rate, $R(x) = x H(x)$ the ramp
function ($H$ the Heaviside step
function), and $b$ a numerical parameter
to control the stiffness. This will allow us to
distinguish the response of the cells to extension ($d^{(+)}$),
compression ($d^{(-)}$), or both ($d^{(\pm)}$). In all cases, when
$d=1$, attachments to the substrate will break, which will lead to a
cell that stops moving, since the propulsion term depends linearly on
$A$, and tries to recover its circular shape.

\begin{figure}[ht!]
  \centering
  \includegraphics[width=0.48\textwidth]{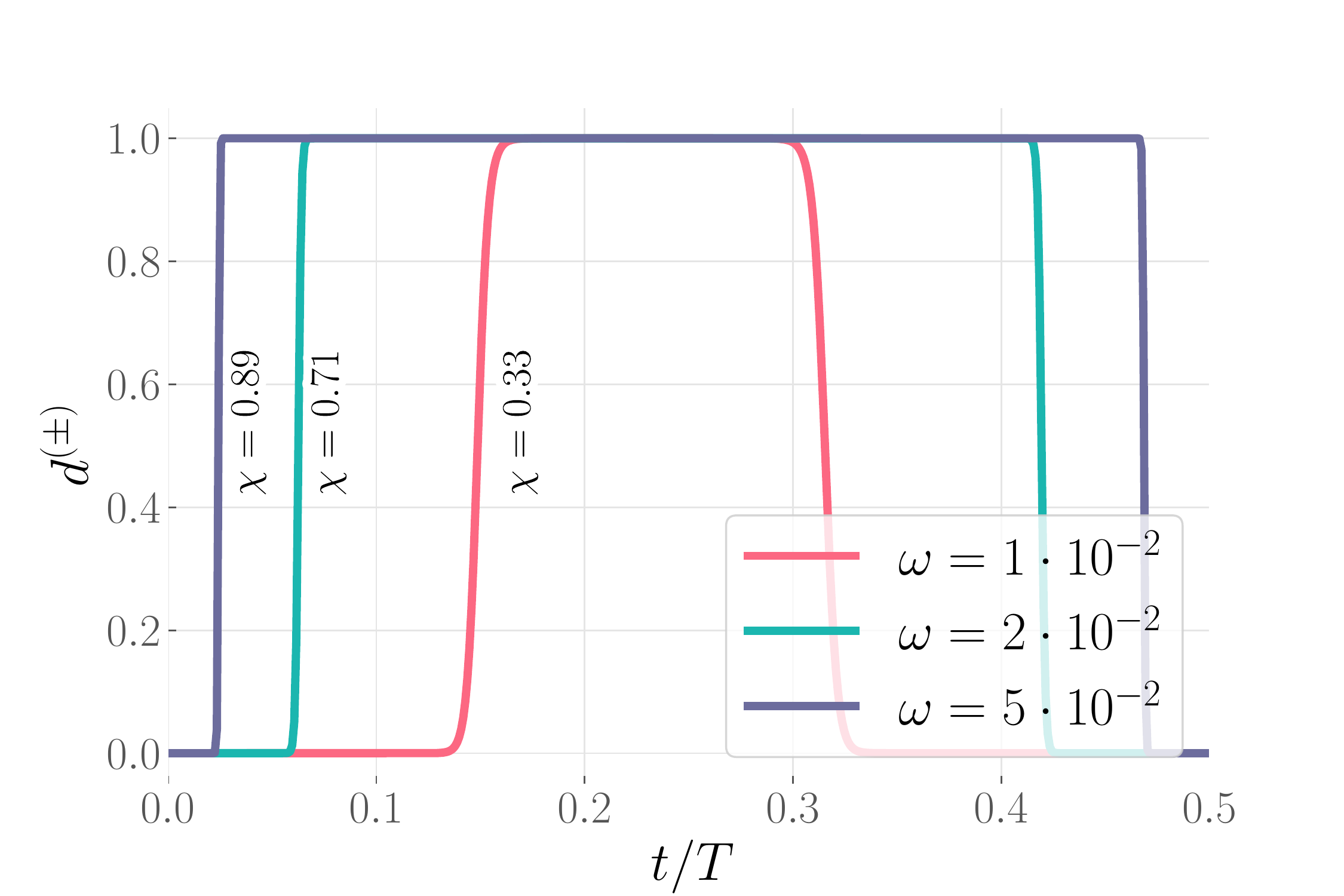}
  \includegraphics[width=0.48\textwidth]{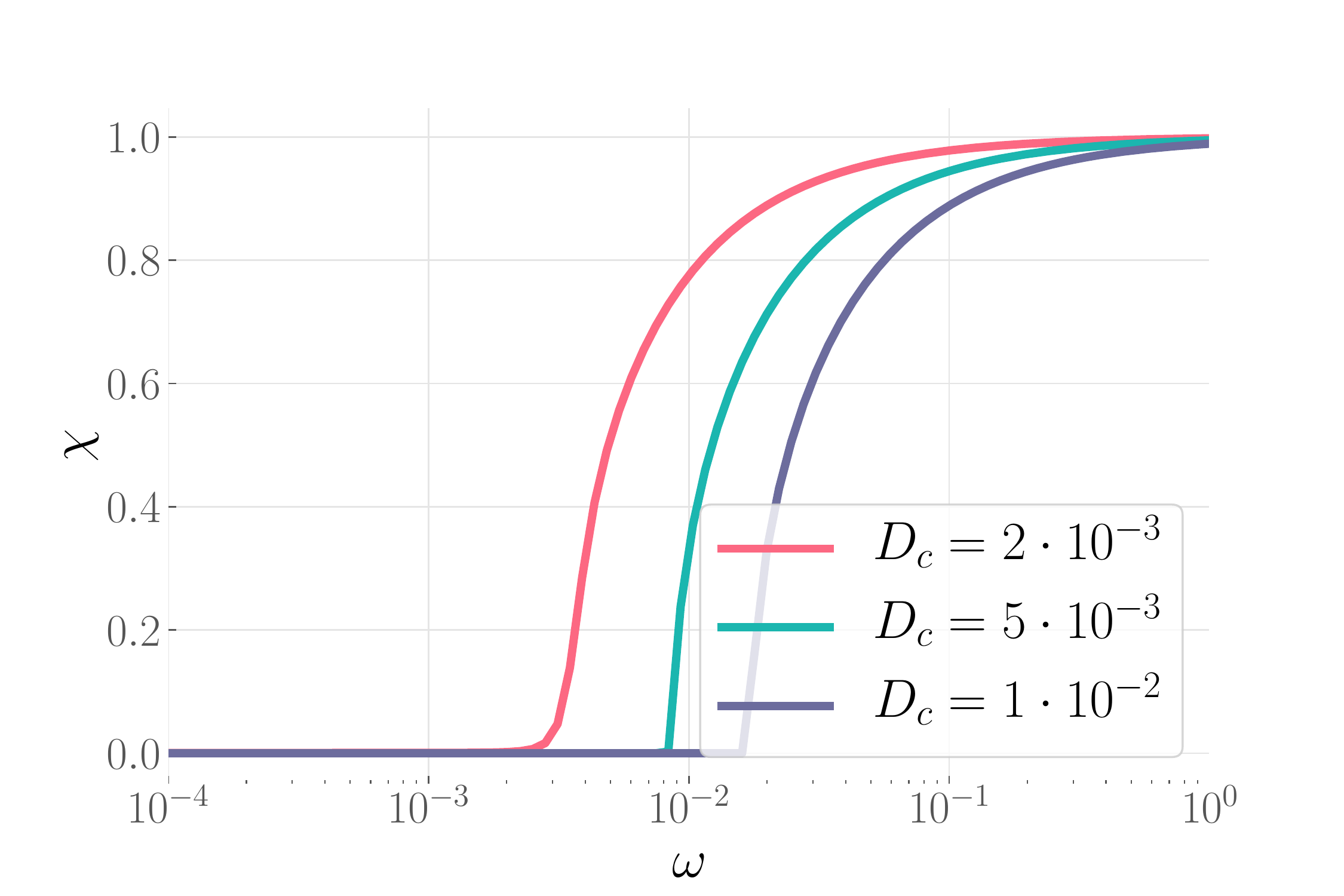}
  \caption{\label{f:chi}(color online) (left) Detachment rate as a function of time
    for three different frequencies, with $D_c = 5\cdot 10^{-3}$,
    $b=10^3$, and $d=1$. (right) Average detachment rate $\chi$, as a function
    of frequency, for three different critical stretching rates $D_c$.}
\end{figure}
\begin{figure}[ht!]
  \centering
  \includegraphics[width=0.7\textwidth]{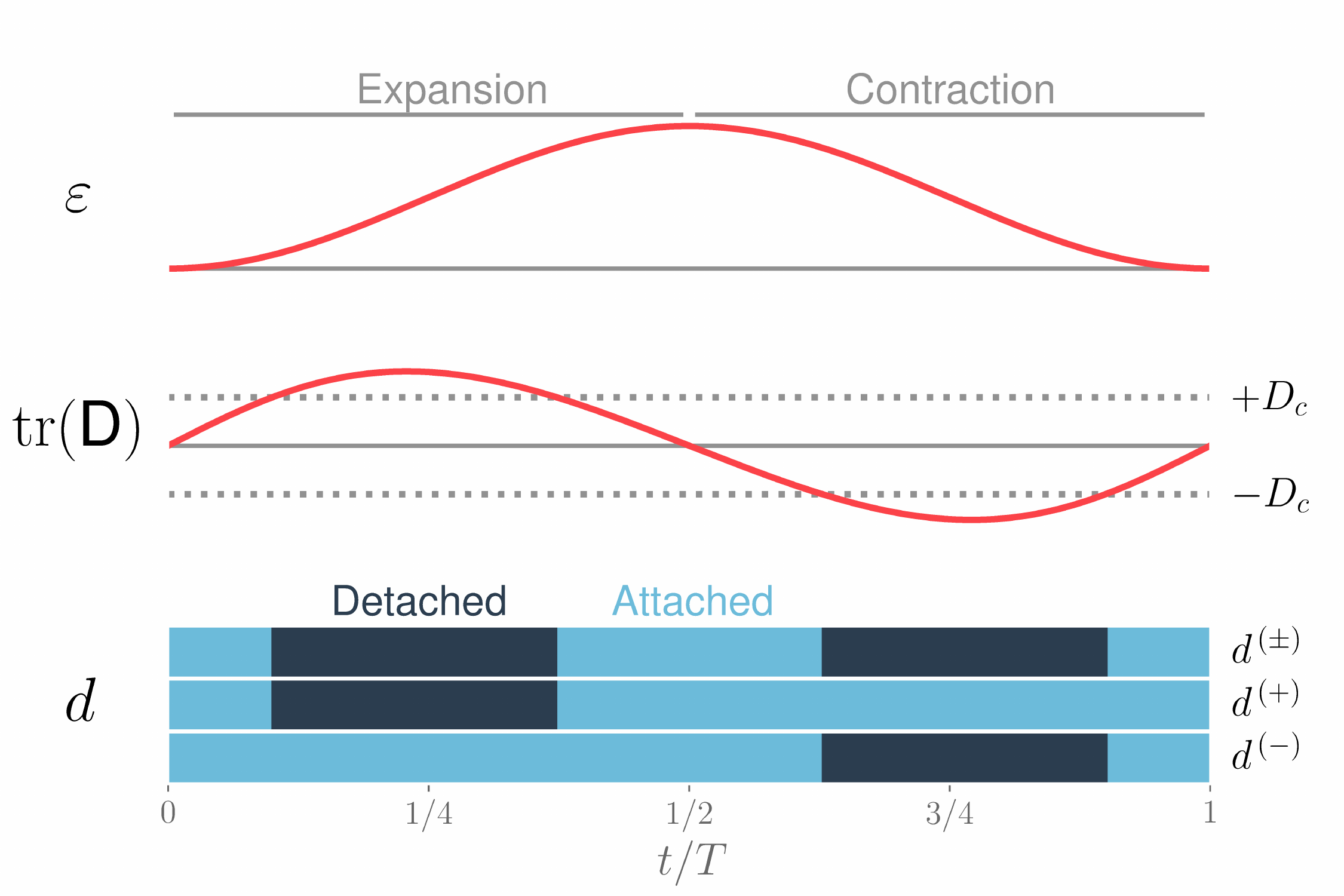}
  \caption{\label{f:eDd}(color online) Schematic representation of the
    adhesion/substrate coupling. From top to bottom, the stretch ratio
  $\eps$, the trace of the rate-of-deformation tensor
    $D=\trace{\parg{\tensor{D}}}$, and the magnitude of the detachment rate
    $d(D)$, with $d=0$ ($d=1$) shown as light and dark blue, respectively, for three different response functions,
    expansion-contraction $d^{(\pm)}$, expansion
    $d^{(+)}$, and contraction $d^{(-)}$.}
\end{figure}

To estimate the critical frequency $\omega_c$, we assume that the
detachment functions (Eqs.(\ref{e:Dpm}-\ref{e:Dm})) exhibit a step-like response, which is a good approximation if $b$ is
large enough (see Figure~\ref{f:chi}). We then have $d=1$ for $D^2 - D_c^2 \ge 0$, which leads to the
following quadratic equation for $y= 1 - \cos{\parg{2\pi\omega t}}$,
from which we can directly compute $\omega_c$ as a function of $D_c$
\begin{align}
  B^2 =
  \frac{y\parg{2-y}}{\parg{1 + \frac{\eps_0}{2} y}^2}
\end{align}
with $B = \frac{D_c}{\pi \omega \eps_0 (1-\nu)}$. The roots to
this equation are given by
\begin{align}
  \parg[\Big]{1 + \parg{B \eps_0/2}^2}\cos{\parg{2\pi\omega t}} &=
  B^2 \frac{\eps_0}{2}\parg[\bigg]{\frac{\eps_0}{2} + 1} \pm \sqrt{1 -
    B^2\parg{1 + \eps_0}}
\end{align}
and are real only if the term inside the square
root is greater than zero, from which we can derive the critical frequency
\begin{align}
  \omega_c = \frac{D_c}{\eps_0 (1-\nu) \pi}\sqrt{1 + \eps_0}\label{e:wc}
\end{align}
Finally, to quantify the degree to which this detachment rate affects
the dynamics, we define a function $\chi$ that measures the average
detachment rate over a half-cycle
\begin{align}
  d_0 \chi &= \frac{2}{T} \int_{t_0}^{t_0 + T/2}
  d (D(t)) \,\df{t}
\end{align}
where $d$ is one of $d^{(\pm)}$, $d^{(+)}$, or $d^{(-)}$.
Alternatively, this also provides a measure of the relative
time-interval during which the cell can move. Fig.~\ref{f:chi} shows the
detachment rate as a function of time, as well as the average
detachment rate as a function of frequency, Fig.~\ref{f:eDd} gives a
schematic diagram of the three main quantities involved in determining
the response of the cell: the time-dependent strain, the rate of
deformation $D$, and the detachment rate $d(D)$.
\section{Simulation and Analysis Method}
\begin{table}[ht!]
  \begin{tabular}{lll}
    \hline
    Parameter     & Value      & Description\\
    \hline
    $\alpha$      & $4$      & Propulsion rate \\
    $\beta$       & $\alpha/2$ & Actin nucleation rate \\
    $\gamma$      & $0.5$      & Motors' symmetry breaking \\
    $\sigma$      & $1.3$      & Motors' contraction \\
    $\mu$         & $0.1$      &Stiffness of volume conservation \\
    $D_\rho$      & $1$        & Stiffness of the diffuse interface \\
    $D_p$         & $0.2$      & Diffusion coefficient for $\boldsymbol{p}$ \\
    $\tau_1^{-1}$ & $0.1$      & Degradation rate of actin \\
    $\tau_2^{-1}$ & $0.4$      & Decay rate of $\boldsymbol{p}$
    outside of cell \\
    $\epsilon$    & $37.25$    & Regularization of actin creation\\
    $D_A$         & $1$        & Diffusion of adhesion sites \\
    $a_0$         & $0.01$     & Linear adhesion attachment rate\\
    $a_{\textrm{nl}}$ & $1.5$  & Nonlinear adhesion attachment rate \\
    $s$           & $1$        & Saturation of adhesion sites \\
    $d_0$           & $1$      & (Maximum) Adhesion detachment rate\\
    $\tau_A^{-1}$ & $\tau_2^{-1}$ & Decay rate of adhesion sites
    outside of cell \\
    $\nu$         & $0.3$      & Poisson ratio \\
    $\omega$      & $0-0.1$    & Substrate Stretching frequency \\
    $\eps_0$      & $0.3$      & Substrate deformation amplitude \\
    $D_c$         & $10^{-3}-10^{-1}$        & Critical rate-of-deformation\\
    $b$           & $10^3$      & Stiffness parameter for detachment
    rate response\\
    $r_0$         & $15$       & Radius of circular initial condition
  \end{tabular}
  \caption{\label{t:params} Default simulation parameters adapted from
    Ref.\cite{Lober2014}.}
\end{table}
We consider a single cell on a cyclically stretched substrate, at
various frequencies, and study the time-dependent orientation for the
three different response functions introduced above $d^{(\pm)}$,
$d^{(+)}$, and $d^{(-)}$. As a reference, we have also considered the
case when $d=0$, as it serves to identify to what degree the
reorientation can be attributed to the passive deformation of the cell
by the substrate. Since we are interested in studying the
frequency dependence of the cell dynamics, we have fixed all
parameters related to the cell and substrate. Unless otherwise stated,
the default values are those listed in Table~\ref{t:params}, which
where taken from a previous study on patterned substrates
performed by Ziebert and Aranson\cite{Lober2014}. In the absence of
stretching, a polarized cell with these parameters will settle into a steady gliding motion
with a fan-like shape. Regarding the stretching protocol, we
follow the experiments of Iwadate et al\cite{Iwadate2009}, and set the Poisson's
ratio at $\nu=0.3$, with a fixed amplitude of $\eps_0 = 0.3$. For all
simulations we considered a single cell of circular radius $r_0 = 15$ that was initially polarized at an angle $\theta_0$ with respect to
the stretching direction ($\theta=0$). In order to quantify the
reorientation response we performed simulations for $n=5$ different
initial conditions $\theta_0 = n \pi/12$ for each set of parameter
values ($\omega$ and $d$).
The initial values for the magnitude of the polarization field and the concentration of adhesion
sites were set to $p = 0.5$ and $A=0.1$, respectively. The dimensions
of the (unstretched) domain were $L_x = L_y = 100$ and we used $N=256$
grid points along each dimension to discretize the system.

To track the orientation of the cells, we computed the center of mass as a
function of time and from this, the (relative) center of mass velocity
within the lab frame was obtained and used to define $\theta$. Specifically, let
$\Delta \bm{r}(t_0, t_1) = \bm{r}(t_1) - \bm{r}(t_0)$ be the center of mass displacement, within the
lab frame, in a time interval $\Delta t = t_1-t_0$. To compute the
relative velocity of the cell $\bm{v}_{\text{eff}}$ with respect to the substrate we should
remove the displacement corresponding to the externally imposed
strain. Consider a substrate element that at time $t=t_0$ coincides
exactly with the position of the center of mass $\bm{r}(t_0)$. The
Lagrangian coordinates for this element are $\bm{\xi}(\bm{r}(t_0),
t_0)$. The spatial position of this
element at any subsequent time $t_1$ is known exactly, since it follows the
substrate deformation, and allows us to define the effective substrate
velocity $\bm{u}(t_0,t_1)$ as
\begin{align}
  \bm{u}(t_0,t_1) &= \frac{\bm{x}\parg{\bm{\xi}(\bm{r}(t_0),t_0), t_1}
    - \bm{r}(t_0)}{\Delta t}
\end{align}
Thus, the effective velocity of the cell, within the lab frame is simply
\begin{align}
  \bm{v}_{\text{eff}}(t_0, t_1) &= \frac{1}{\Delta t}\barg[\Big]{\Delta
    \bm{r}(t_0, t_1) - \Delta t \bm{u}(t_0, t_1)}
\end{align}
To obtain accurate measurements for $\bm{v}_{\text{eff}}$ we made sure
that the sampling time $\Delta t$ was smaller than both the period of
oscillation ($T=\omega^{-1}$)
and the time $\tau$ required for the cell to move a distance
equal to its diameter in the static case ($\omega=0$), with $v_0$ the
steady state velocity, such that $\tau = r_0/v_0$.

The shape deformations of the cells can be tracked by computing
the aspect ratio $h$, defined in terms of the
following shape tensor \cite{Ziebert2012}
\begin{align}
  I^{ij}&= \int (x^i - R^i)(x^j - R^j) \df{x}\df{y}
\end{align}
where $\bm{R} = \avg{\bm{x} \rho(\bm{x})}$ is the center of mass position
of the cell. The aspect ratio is then given as $h =
\sqrt{\lambda_1/\lambda_2}$, where $\lambda_1$ and $\lambda_2$ are the
eigenvalues of $I$ ($\lambda_1\ge \lambda_2$). A cell in the circular (static) state will have an aspect ratio of
$h=1$, whereas the fan-like crawling cells in the absence of stretching will present an aspect ratio closer to $h=2$.

\begin{figure}[ht!]
  \centering
  \includegraphics[width=0.9\textwidth]{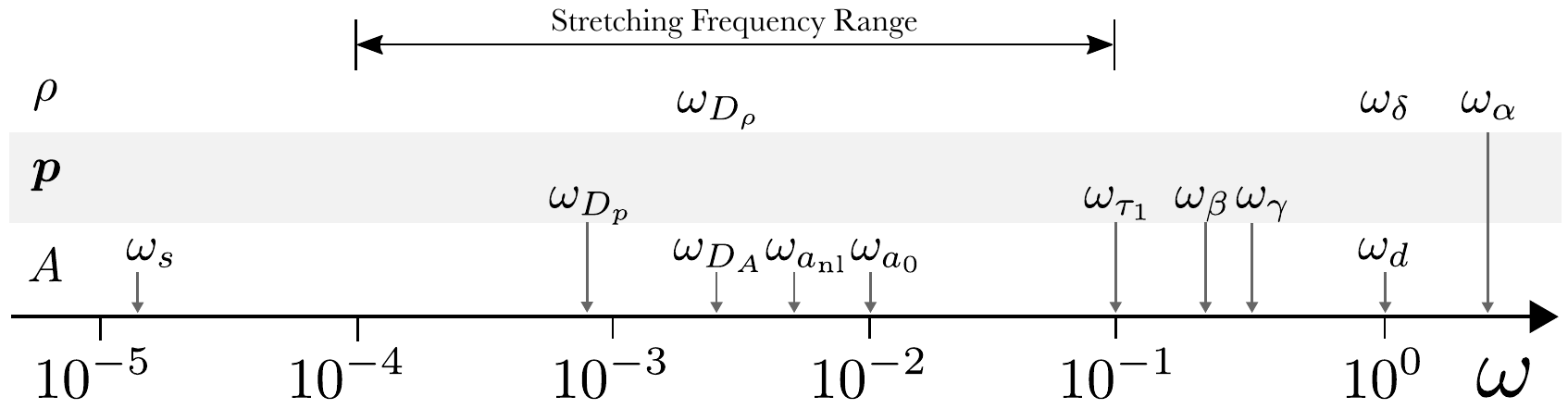}
  \caption{\label{f:frequencies}Relevant time/frequency scales for the
  cell $\rho$, polarization $\bm{p}$, and adhesion $A$ dynamics for
  the default choice of parameter values given in
  Table~\ref{t:params}. The stretching frequency range is chosen to
  probe the role of the adhesion dynamics on the cell response.}
\end{figure}
To understand the reorientation dynamics, we need to consider the interplay between
the dynamics of the
shape deformations, the actin dynamics, and the adhesion dynamics, as
well as their characteristic time-scales, and how they compare to the
time-scale over which the substrate is being deformed. For this we
first define the characteristic length scales in the system, the cell
size $r_0 = 15$ and the interface thickness $\zeta = D_\rho^{1/2} = 1$. The
characteristic time or frequency of the shape deformations is determined
by the stiffness of the interface as $\omega_{D_\rho} = D_\rho r_0^{-2}
\simeq 4\cdot 10^{-3}$, as well as the time governing the
retraction/expansion of the two phases $\omega_\delta = 1$. The frequency associated to the propulsion of
the cell by the actin network is $\omega_\alpha = \alpha \zeta^{-3} \simeq 4$. The time-scales for the actin dynamics include the diffusion
time-scale ($D_p$), the depolymerization rate ($\tau_1$), the
polymerization rate ($\beta$), and the asymmetry driving term
($\gamma$). They in turn yield the following characteristic
frequencies, $\omega_{D_p} = D_p r_0^{-2} \simeq 9\cdot 10^{-4}$,
$\omega_{\tau_1} \simeq \tau_1^{-1} = 10^{-1}$, $\omega_\beta =
\beta \epsilon^{-1/2} \simeq 3 \cdot 10^{-1}$, and
$\omega_\gamma = \gamma \zeta^{-1} \simeq 5\cdot 10^{-1}$. Finally, the
frequencies associated to the adhesion dynamics are $\omega_{D_A} =
D_A r_0^{-2} \simeq 4\cdot 10^{-3}$, $\omega_{a_0} = a_0
\zeta^2 = 10^{-2}$, $\omega_{a_{\text{nl}}} = a_{\text{nl}}
r_0^{-2} \simeq 7\cdot 10^{-3}$, $\omega_s = s r_0^{-4} \simeq
2\cdot 10^{-5}$, and $\omega_{d} = d = 1$. Where it should be noted
that the relevant length scale for the linear attachment rate for the
adhesions ($a_0$) is the
characteristic size of the region over which the $\bm{p}$ field is non-zero. Analysis of the simulations shows that this is strongly peaked
near the leading edge, as the polymerization rate is proportional to
$\grad\rho$. For simplicity we have assumed that this is given by the
interface width $\zeta$, but this is just a lower bound, the real
value should be slightly higher $\simeq 2\sim 4\zeta$. In contrast, the
relevant length-scale for the non-linear growth and
saturation rates is the cell size $r_0$. To summarize, using the
default parameter values given in Table~\ref{t:params}, we can identify the
following frequency regimes governing the shape ($\rho$), actin
($\bm{p}$), and adhesion ($A$) dynamics
\begin{align*}
  \omega_{D_\rho} &\ll \omega_\delta < \omega_\alpha\\
  \omega_{D_p} \ll \omega_\gamma &< \omega_{\tau_1} <
  \omega_\beta \\
  \omega_s \ll \omega_{D_A} &< \omega_{a_{\text{nl}}} <
  \omega_{a_0} \ll \omega_d
\end{align*}
An illustration of the different characteristic frequencies is given in
Fig.~\ref{f:frequencies}. Here, since we are interested in studying
how the adhesion dynamics affect the reorientation, we will focus on
stretching frequencies within the range $10^{-4} < \omega < 10^{-1}$,
such that  $\omega_{D_p} < \omega < \omega_{a_0}$.
\section{Results}
\label{s:res}
\subsection{Passive alignment}
\begin{figure}[ht!]
  \centering
  \includegraphics[width=0.35\textwidth]{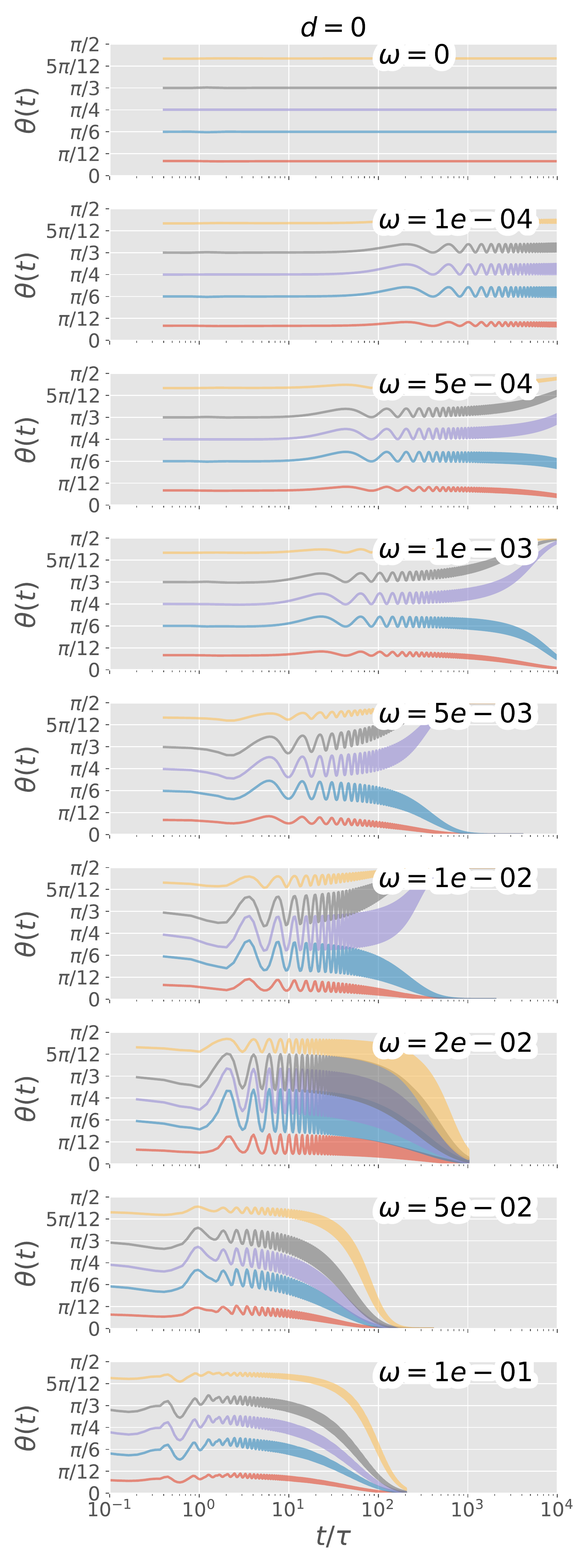}
  \caption{\label{f:passive} (color online) Orientation $\theta$ as a function of
    time $t$ for different initial polarization directions $\theta_0$,
    and various frequencies, in the absence of any
    specific cell-substrate coupling ($d=0$). Default parameters given
    in Table~\ref{t:params} were used. Note that $\theta=0$
    ($\theta=\pi/2$) corresponds to parallel (perpendicular) alignment.}
\end{figure}
Let us start by considering the simple case of a cell that is being
passively advected by the substrate, in the absence of any direct
coupling, i.e., $d=0$. The results for this case
are summarized in Fig.~\ref{f:passive}, which shows the orientation as
a function of time, for various stretching frequencies $\omega$. 
First, in the absence of stretching, for $\omega=0$, the orientation of the cell is time-independent, as
expected, since we have not included any source of stochasticity in
the model. For non-zero frequencies, the orientation shows a clear
time-variation, since it is being constantly deformed and rotated by
the substrate. However, there are several distinct frequency regimes,
depending on how the stretching frequency compares to the
characteristic frequencies of the system,
giving rise to qualitatively different realignment dynamics. At very low frequencies, $\omega \lesssim 10^{-4}$, the
orientation oscillates around the initial value, but there is no
stretch-induced alignment. In this case, the deformation is so slow
that the cell (together with the actin network and adhesion sites) can completely follow the imposed strain. As the
frequency is increased further, such that it becomes comparable to the
frequency for the diffusion of orientational order $\omega_{D_p}$, we
begin to see an alignment either parallel or perpendicular to the
stretching direction. In this case, the actin network is not able to
rearrange fast enough to adapt to the changing shape of the
cell. However, this alignment is extremely slow, with a time-scale of the order of $t/\tau\simeq 10^{4}$. In addition, there seems to be
no preference between parallel or perpendicular directions, with the
final orientation depending on the initial
orientation: cells that were aligned closer to the parallel or
perpendicular directions will favor those orientations. Upon
increasing the frequency of oscillation to $\omega \simeq 5\cdot
10^{-3}$, the qualitative behavior remains unchanged, but the
reorientation time-scale is reduced by roughly an order
of magnitude. At these frequencies, the substrate is stretching faster
than the cell can relax, since $\omega > \omega_{D_\rho}$, so that the
shape starts to become perturbed by the imposed strain. If the frequency is increased still further, we observe a clear
transition, at $\omega \simeq 2\cdot 10^{-2}$, above which all cells show a parallel alignment, regardless of the initial
orientation. For such high frequencies, $\omega >
\omega_{a_0} \gtrsim \omega_{a_{\text{nl}}}$,  the distribution of
adhesion bonds inside the cell can no longer be stabilized fast enough
to keep track of the imposed deformations.

\begin{figure}[ht!]
  \centering
  \includegraphics[width=0.8\textwidth]{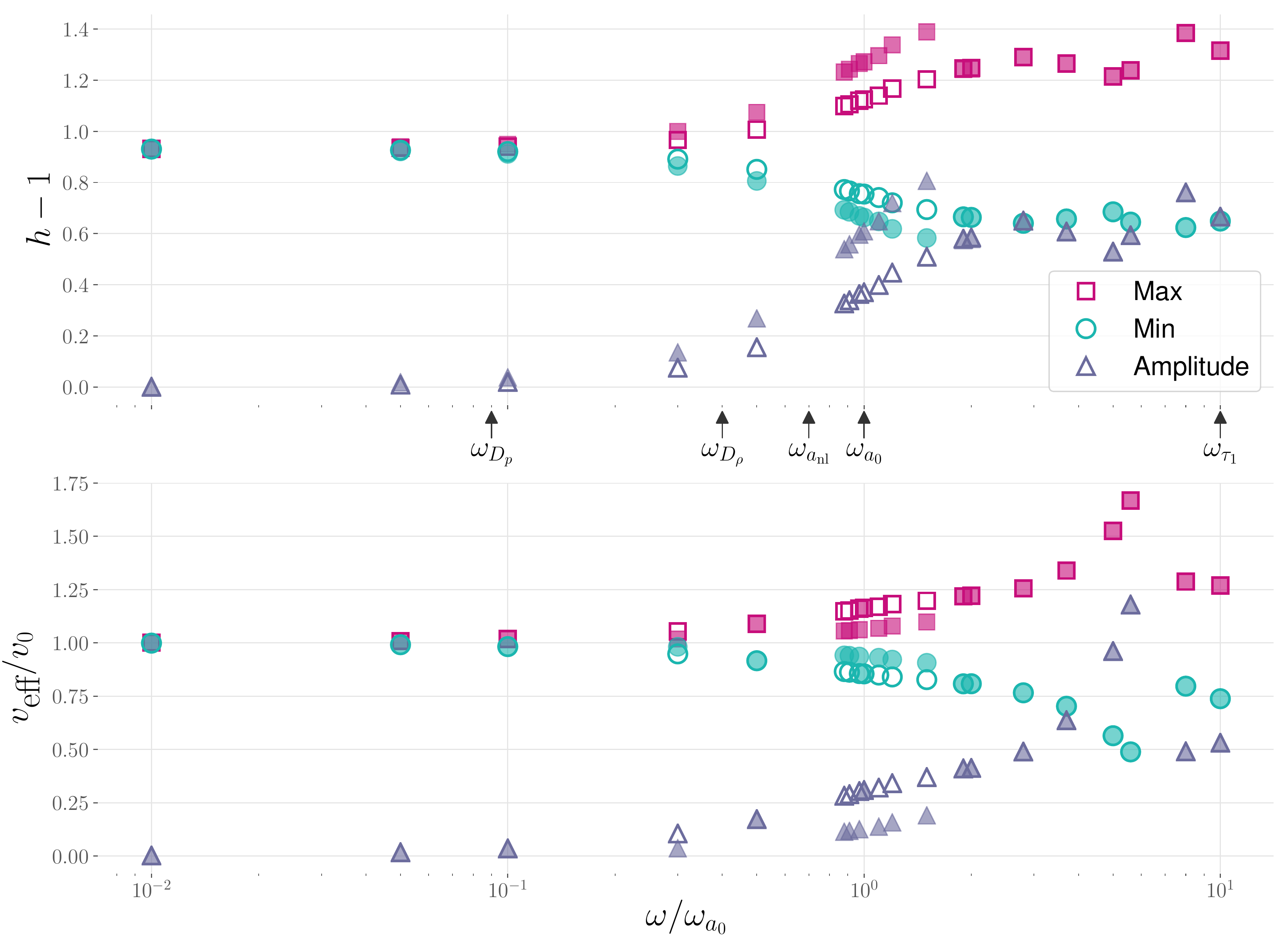}
  \caption{\label{f:passive_hv} (color online) Maximum, minimum, and
    amplitude of the (steady) oscillations of the aspect ratio $h$ and
    effective velocity $v_{\text{eff}}$ as a function of
    frequency $\omega$. While there is
    no significant change in the average values of $h$ or
    $v_{\text{eff}}$, the fluctuations of these quantities depend
    strongly on frequency. Data was obtained from the trajectories
    of cells initially polarized in the $\theta=\pi/6$ (open symbols)
    and $\theta=\pi/3$ (filled symbols) orientations, once the
    orientation of the cell was stabilized. The amplitude was computed
    as the difference between the maximum and minimum values.}
\end{figure}

As a complement to the previous analysis, we
can also consider the time-dependence of the aspect ratio $h$ and the
magnitude of the effective cell velocity $v_{eff}$. The time-variation
of these quantities shows similar oscillations in response to the
strain as does the orientation $\theta(t)$, but there is no systematic
drift, with both quantities oscillating around their ``equilibrium''
($\omega = 0$) values, corresponding to $h_0\simeq 1.9$ and
$v_0\simeq 0.6$. Studying how the fluctuations in these
quantities changes as a function of frequency will help us to clarify
the mechanosensitive response of the cells. For this, we have plotted
the maximum and minimum value of $h-1$ and $v_{0}$, as well as the
amplitude of the corresponding oscillations, for two different initial
orientations ($\theta = \pi/6$ and $\pi/3$) in
Figure~\ref{f:passive_hv}. As expected, at lower frequencies $\omega <
\omega_{D_{\rho}}$ the fluctuations are negligible, as the cell is
able to relax to its preferred shape faster than the substrate is
being deformed. In addition, even though the cells reorient into
either the parallel or perpendicular directions for $\omega \gtrsim
\omega_{D_p}$, we see no difference in their shape or velocity. This
means that for this frequency range the reorientation of the cell can
be effectively decoupled from its translational motion. As the
frequency becomes comparable to $\omega_{D_\rho}$ the shape of the
cell begins to show oscillations, as the substrate is moving faster
than it can relax. Since the velocity and motility are intimately
linked, this is accompanied by a corresponding increase in the
velocity fluctuations, but this effect is much less pronounced. It is
at this point where we can start to see a difference between cells oriented
perpendicular or parallel to the stretching. The cell that was
initially polarized at $\theta = \pi/3$ will align in the
perpendicular direction and experiences considerable shape deformation
but relatively small velocity fluctuations. The cell that was
polarized in the $\theta = \pi/6$ direction will align in the parallel
direction and shows the opposite behavior, small shape deformations but
large velocity fluctuations. These tendencies increase with increasing
frequency, up until $\omega \simeq \omega_{a_0}$, where the only
stable orientation is the parallel one. Here, the fluctuations of the aspect
ratio reach a plateau, which tells us that the cell shape is now completely
unable to respond to the imposed strain. Simultaneously, at this point the
stretching starts to interfere with the adhesion dynamics and this
greatly amplifies the fluctuations in the velocity. We see that the
cell can slow down and speed up by up to $50\%$ with respect to their average
value. This is due to the heterogeneous and unstable distribution of
adhesion sites that characterize the cell at these frequencies. These fluctuations
reach a maximum at $\omega \simeq 5\omega_{a_0}$, after which their
amplitude shows a sharp decrease. 

The question of why the cells choose one particular orientation
over another, and why the only stable orientation is the parallel one
at high frequencies remains to be answered. Existing theories\cite{De2007,De2008,Safran2009,Livne2014a,Xu2016}, which focus on
slow crawling cells with stress fibers, and do not consider shape
deformations or the cell motility, predict that both parallel $\theta=0$ and
perpendicular $\theta = \pi/2$ orientations are solutions to the
steady state equation ($\text{d}\theta/\text{d}t = 0$), together with
an oblique orientation $\theta_f$, which is a function of the system
parameters. While the oblique (nearly perpendicular) orientation
$\theta_f$ is usually the stable solution, under certain conditions, such as when the
mechanical forces due to the substrate dominate the cellular
activity, or if the substrate is very soft, the parallel orientation becomes stable\cite{De2007,De2008,Xu2016}. A direct comparison with our
results is not straightforward, but we also find $\theta=0$ and
$\theta=\pi/2$ as steady state solutions, with the parallel
orientation the only stable one at high frequencies (at least within
the frequency range we have considered). In this high frequency regime,
we have seen that the cell is unable to resist the shape deformations
imposed by the substrate, and that the distribution of adhesions is
unstable, leading to large velocity fluctuations, even though the
average velocity remains unchanged. In this limit, the
forces due to the externally imposed strain dominate any forces due to
the intrinsic cell motility. Thus, our findings of a stable parallel
orientation are consistent with the theoretical predictions\cite{De2007,De2008,Xu2016}.

In the absence of any specific cell-substrate interaction, the
strong coupling that exists between shape and motility yields a
preferential alignment under cyclic stretching that is strongly
dependent on the relative frequency. At very low frequencies the
cells and the actin network have time to readjust to the deformation,
and the average migration direction is not affected. When the
stretching is faster than the actin network can respond, there is a
very weak reorientation process, but no preference between
perpendicular or parallel directions. If the stretching is
faster than the time-scale over which the cell can accommodate its
shape (as defined by the stiffness of the membrane), then the
reorientation is significantly faster. At even higher frequencies,
where the cell cannot form and stabilize the adhesion bonds fast enough to follow
the deformation, we observe that the cells align parallel to
the stretching direction, regardless of initial orientation. Thus,
even without any direct coupling to the
substrate, there is a clear preference in the direction of migration.
\subsection{Active alignment}
\begin{figure}[ht!]
  \centering
  \includegraphics[width=0.9\textwidth]{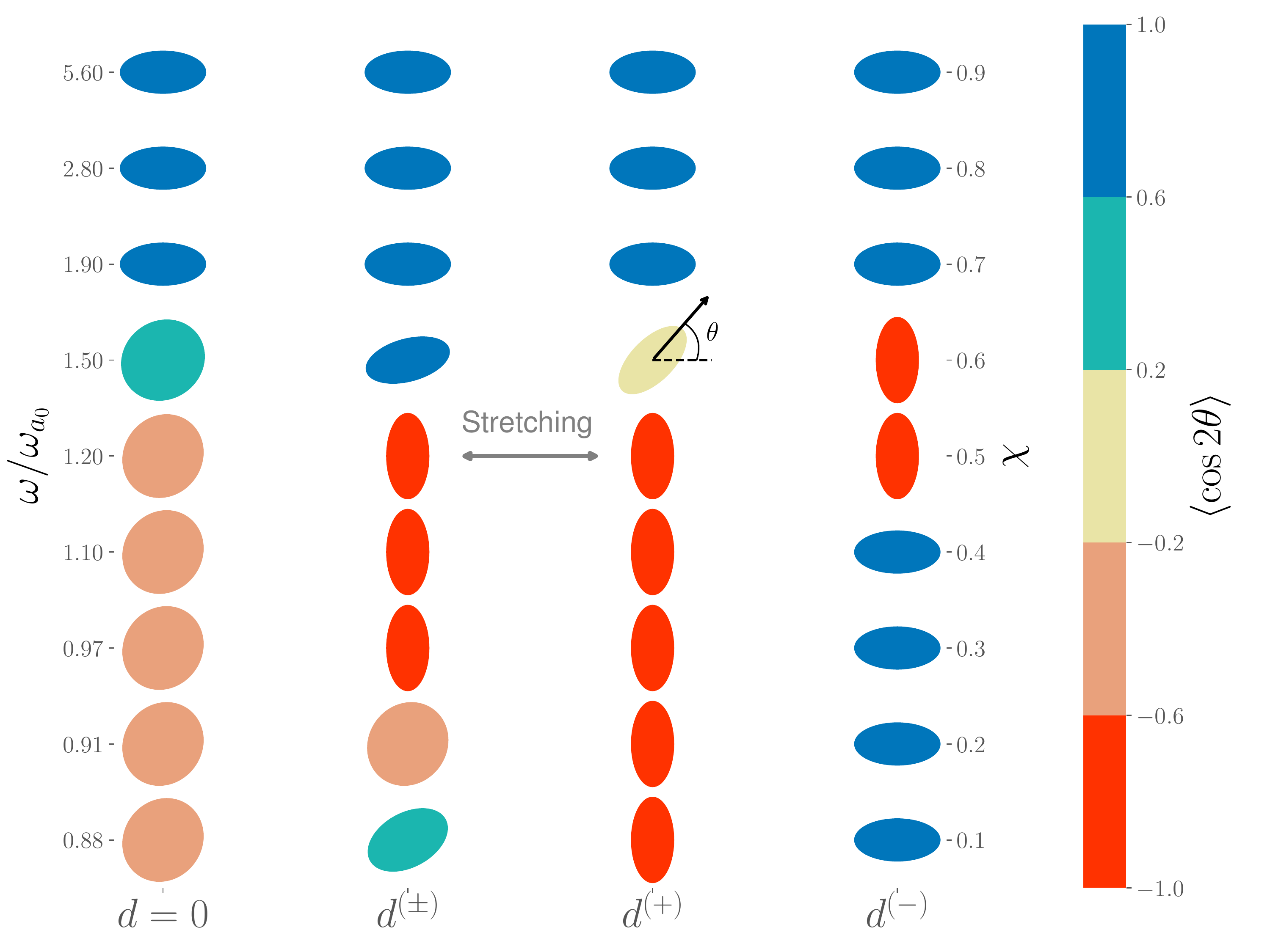}
  \caption{\label{f:active}(color online) Phase diagram showing the final orientation
    of the cells as a function of frequency $\omega$ (average detachment rate
    $\chi$) and adhesion response function. Each point is specified as
    an ellipse, with orientation, aspect ratio, and color used to encode
    information on the average orientation and the spread of the
    orientations. Results
    from five simulations with distinct initial cell polarization
    directions ($\theta_0 = \pi/12, \pi/6, \pi/4, \pi/3, 5\pi/12$)
    are used for each point. The average ($\langle\theta\rangle$) and
    the standard deviation $\big(\sqrt{\langle\theta^2\rangle}\big)$ of
    the steady-state orientation is used to define the orientation of
    the long axis and the aspect ratio of the ellipse,
    respectively. In this case, small (large) standard deviations
    result in elongated (spherical) shapes. Finally, we also compute an orientational order parameter $\langle\cos(2\theta)\rangle$, which is $1$ ($-1$)
    if all cells align parallel (perpendicular) to the
    stretching, and use it to color-code each ellipse.
    The critical stretching rate was set by $D_c = 5\cdot 10^{-3}$,
    all other parameters are the same as in Fig.~\ref{f:passive}.}
\end{figure}
We now consider the reorientation of cells whose internal propulsion
mechanism is actively responding to the strain it receives from the
substrate through a rate-dependent detachment rate ($d\ne 0$). This is done to
model the frequency dependent stability of the focal adhesions\cite{Kong2008,Zhong2011} As
described above, Eqs.(\ref{e:Dpm}-\ref{e:Dm}), we will consider cells
that respond to either compression $d^{(-)}$ or extension $d^{(+)}$,
or both $d^{(\pm)}$. By setting the threshold value $D_c$ ($\omega_c$)
at which this response is activated, we can control the interval during which the cell is able
to form attachments, and thus crawl over the substrate. Taking into
account the results presented above in the absence of any direct
coupling $d=0$, for which the cells show parallel reorientation when
the frequency of oscillation is greater than the frequency associated
to the attachment dynamics, we can expect that a rate-dependent detachment rate will
significantly affect the reorientation dynamics. We set $D_c =
5\cdot 10^{-3}$ ($\omega_c \simeq 8\cdot 10^{-3} \lesssim
\omega_{a_0}$) and the frequency to lie in the range of $8\cdot 10^{-3} < \omega <
6\cdot10^{-3}$ ($\chi$ between $0$ and $1$). Within this
range, the only relevant time scales are those corresponding
to the attachment dynamics $\omega_{a_\text{nl}}$ and $\omega_{a_0}$,
as the time scales for the actin and shape deformations are
both slower $\omega_{D_p} < \omega_{D_\rho} < \omega$. For these lower
frequencies, we would have $d=0$ and the dynamics would be the same as in
the passive case (Fig.~\ref{f:passive}). We have summarized
the results in the phase diagram presented in Fig.~\ref{f:active} (see
ESI~1 for the full set of trajectory data). For comparison purposes,
the corresponding results for $d=0$ have also been included. First,
at high frequencies ($\omega\gtrsim 2\cdot 10^{-2}$), we
see that all cell types show parallel alignment, regardless of the specific form of the
response function. In such cases, the stretching is too fast for the
cell to respond ($\omega > \omega_{a_0} > \omega_\rho > \omega_p$), so the
exact details of the attachment/detachment become irrelevant. More interesting are the
results at low and intermediate frequencies. At low frequencies
$8\cdot 10^{-3}\lesssim \omega \lesssim 1.1\cdot 10^{-2}$ ($0.1\le
\chi \le 0.4$), cells with
$d^{(+)}$, which ``resist'' extension, exhibit a perpendicular
alignment. In contrast, cells with $d^{(-)}$, which ``resist''
compression, show a parallel alignment. Within this frequency range the
cyclic detachments occur over time-scales comparable to the time it
takes for the cell to form and grow new attachments. It is clear that this alignment is due to the type of
detachment, since cells with $d=0$ show no preferential
alignment, with $\theta = 0$ or $\theta=\pi/2$ equally likely. Furthermore, the realignment of
the cells with non-zero detachment rate occurs over time-scales that
are considerably shorter that those with $d=0$. 

\begin{figure}[ht!]
  \centering
  \includegraphics[width=\textwidth]{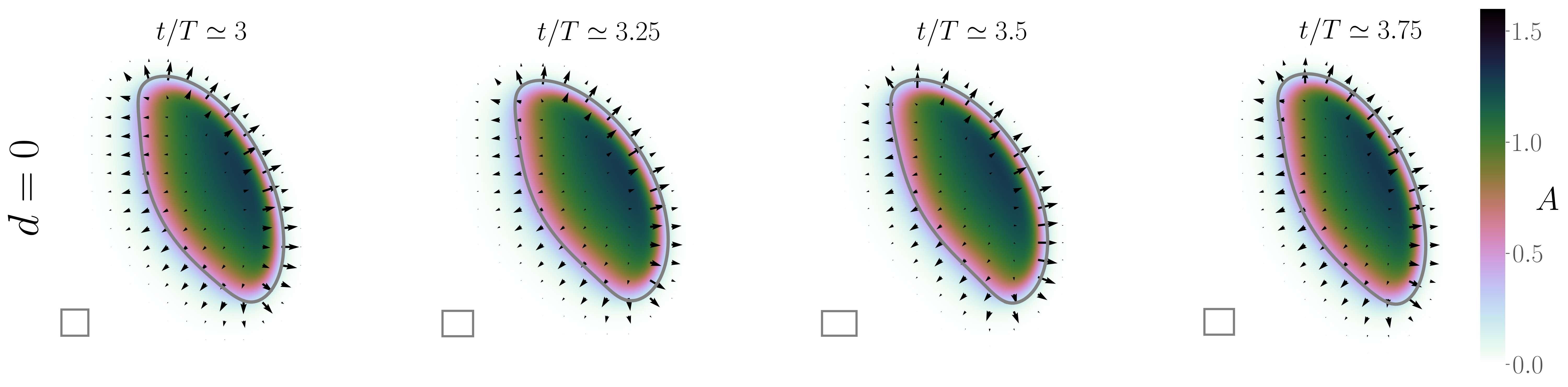}
  \includegraphics[width=\textwidth]{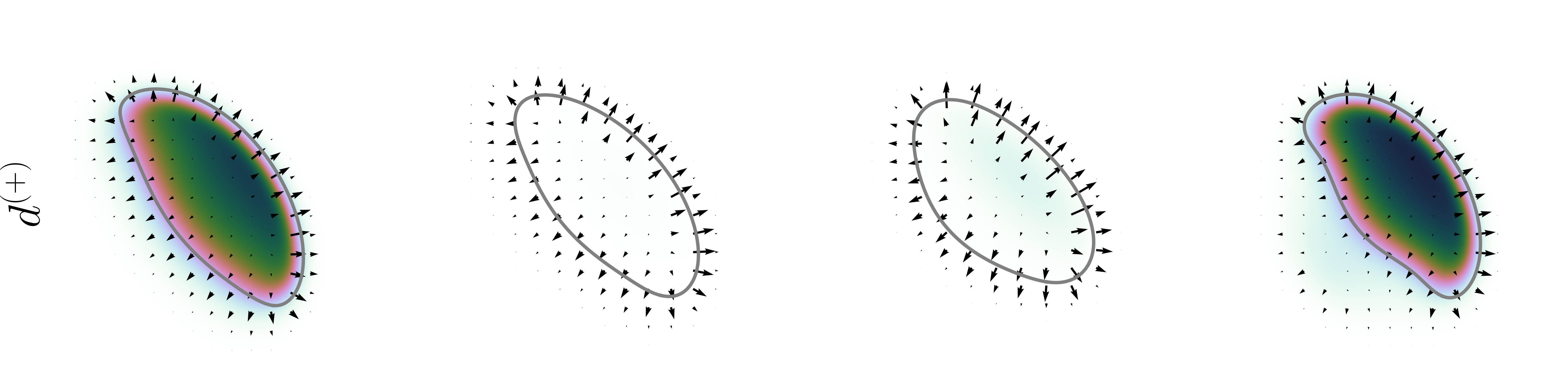}
  \includegraphics[width=\textwidth]{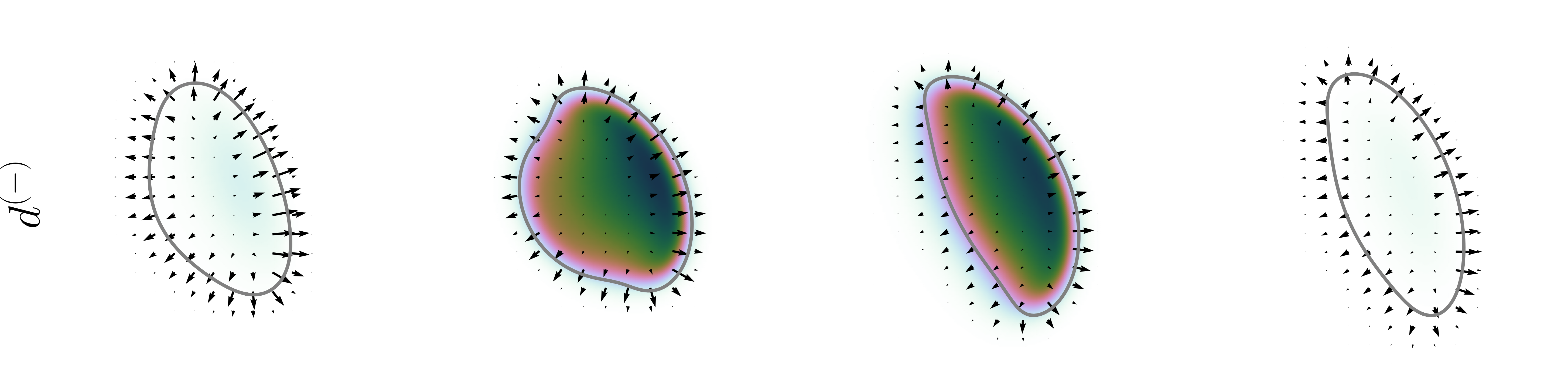}
  \caption{\label{f:snapshots}Simulations snapshots for cells
    exhibiting passive and active responses to a
    periodically stretching substrate. Data was taken over one cycle
    between $3\le t/T < 4$, for cells initially polarized in the
    $\theta = \pi/6$ direction. From top to bottom, $d=0$, $d^{(+)}$, and
    $d^{(-)}$, respectively. Solid lines show the contour of the
    phase-field for $\rho = 0.5$, the density map shows the
    concentration of adhesion sites $A$, and the arrows the actin
    orientation field $\bm{p}$. The rectangles in the top figure show
    the substrate deformation, scaled down by a factor of 20.}
\end{figure}

Simulation snapshots for $\omega=8.8\cdot 10^{-3}$ ($\chi=0.1$), showing the cell shape,
concentration of adhesion sites, and actin orientation are given
Fig.~\ref{f:snapshots}. Compared to cells with $d=0$, cells showing an
active response to the stretching ($d^{+}$ or $d^{-}$) exhibit more
pronounced shape deformations. As can be seen from the figure, the
$d^{(+)}$ cells completely detach as the substrate is extending and
rotating them towards the parallel direction. However, they are able
to recover their adhesions during the compression stage, when they are
being rotated into the perpendicular orientation. Cells with $d^{(-)}$
show the opposite behavior. This asymmetry in the dynamics during the extension and
compression stages is the cause of the reorientation. The corresponding movies are provided as Supplemental
Material (ESI~2-4). For comparison purposes, movies obtained from
simulations at high frequencies $\omega = 2.8\cdot 10^{-2}$ ($\chi =
0.8$), for which all cells show parallel alignment, are also available (ESI~5-7).

For intermediate
frequencies $\omega\simeq \omega_{a_0}$, with $\chi \simeq 0.5$,
$d^{(+)}$ cells exhibit a transition
between the low frequency response (favoring perpendicular
orientations) and the high frequency response (favoring parallel
orientations), resulting in a stable oblique orientation $\theta\simeq
\pi/4$. Surprisingly, the $d^{(-)}$ cells also exhibit a non-monotonic
behavior, even though the low and high frequency limits both show parallel
orientations, at $\omega\simeq 1.2\cdot 10^{-2}$ ($\chi\simeq 0.5$) cells align perpendicular to the
stretching direction. The behavior of the $d^{(\pm)}$ cells seems more
complicated, but it can roughly be understood as a competition between
the
opposing tendencies of $d^{(+)}$ and $d^{(-)}$ cells to orient
perpendicular or parallel to the stretching at low frequencies, with
the perpendicular response being dominant. This is consistent with the
fact that the reorientation time-scale is much longer than that of the
other cell types.

We have performed simulations for two other critical
stretching rates, $D_c=2\cdot 10^{-3}$ ($\omega_c \simeq 3\cdot
10^{-3}$) and $D_c=10^{-2}$ ($\omega_c \simeq 10^{-2}$), and found
similar behavior, at least in the high frequency range. However, for
$\omega_c=3\cdot 10^{-3} \lesssim \omega_{D_\rho} = 4\cdot 10^{-3}$,
the stretching is now able to probe the shape-deformations. For the
lowest frequency considered, $\omega = 3.4\cdot 10^{-3}$ ($\chi=0.1$),
the $d^{(+)}$ ($d^{(-)}$) cells actually favor a parallel
(perpendicular) alignment. Increasing the frequency to $\omega
=3.7\cdot 10^{-3} \simeq \omega_{D_\rho}$ the system reverts back to
being dominated by the adhesion dynamics, thus, if the frequency is
not too high $d^{(+)}$ ($d^{(-)}$) cells will tend to align perpendicular
(parallel) to the direction of stretching. We note however that for
$d^{(-)}$ cells the reorientation response is less pronounced,
particularly at intermediate frequencies. Again, in the high-frequency
range $\omega > \omega_{a_0}$ all cells show parallel alignment.

Experimentally, the fast-crawling cells that we are modeling, such as \textit{Dictyostelium} and
HL-60\cite{Iwadate2009,Iwadate2013,Okimura2016a}, have been shown to
align perpendicular to the stretching direction, just as our cells
within an appropriate frequency range. In these experiments, the imposed strain
was not sinusoidal in nature, but more saw-tooth like: a quick
expansion of the substrate was followed by a static interval and then a
slow relaxation to the original shape (such that the duty
ratio was fixed to $1:1$). Thus, there is a clear asymmetry in the rate of
deformation imposed in experiments during the expansion and
contraction phases. Assuming this is enough to cause a relative instability in the
adhesions during expansion/contraction, it would correspond to our
simulations for $d^{(+)}$. Note that the cells would not need to
be able to distinguish between expansion or contraction (as we have
assumed for our simulations), but only the rate of deformation, since
this rate is different in the two intervals. Our simulations
then provide evidence to favor the adhesion dynamics as being responsible
for the reorientation. This could be tested by using
a reciprocal deformation to that of the original
experiments\cite{Iwadate2009}, with a slow expansion followed by fast
contraction, for which our model ($d^{(-)}$) predicts a parallel
orientation. Finally, while we do not claim quantitative agreement
is possible with the simple model we have used here, particularly
because it has not been parametrized for any specific cell type, we
predict that in the limit where the adhesion dynamics dominates the
response of the cells, an asymmetry in the expansion/contraction
periods of the stretching can be used to selectively drive the reorientation.

\section{Discussion}
\label{s:conc}
The question of cellular realignment under a cyclically
stretching substrate has attracted much attention recently due
to its biological significance. Among the various possible factors or mechanisms
determining this mechanosensitive response, two have been singled out: (1) the viscoelasticity of the actin filament networks and (2) the
focal adhesion dynamics. This is understandable, as the former is
largely responsible for the mechanical properties of the cell, and the
latter provides the coupling between the cell and the substrate
(through the actin-network) needed for the transfer of forces. While
there has been considerable success in developing theories that can
predict the reorientation dynamics of cells under cyclic stretching of
the substrate, several issues remain. First, the exact mechanism
responsible for the reorientation remains illusive. For example, Livne et al.\cite{Livne2014a} attribute it to the passively stored
elastic energy, while Chen
et al. attribute it to the forces on the focal
adhesions\cite{Chen2015a,Qian2013,Xu2016,Xu2018}. Both theories are able to explain the same
set of experimental observations equally well, and even result in the
same theoretical prediction for the orientational dynamics, making it difficult to
determine which of the two effects is the dominant one. Second, most
theoretical and simulation work has so far focused on slow crawling
cells which contain stress fibers, such as fibroblasts. These type of
cells usually align in such a way that their stress fibers are aligned
perpendicular to the stretching direction. Furthermore, since they
move so slowly, their motion can be decoupled from their
reorientation. Therefore, the question of how fast-crawling cells
without stress fibers, such as \text{Dictyostelium}, reorient under
cyclic stretching has remained largely unanswered. Recent experiments by Iwadate et
al.\cite{Iwadate2009,Iwadate2013,Okimura2016a,Okimura2016} have shown
that they prefer to orient perpendicular to the stretching
direction. This perpendicular reorientation is observed without any
corresponding alignment of the dense actin network inside of the
cells.

To study how fast crawling cells respond to large amplitude cyclic
deformations, we require a model that describes both the cell motion
and its reorientation. For this, we need to take into account the
internal machinery of the cell (e.g., the actin-network and myosin
contractility), its coupling to the substrate (through the focal
adhesion sites), and the accompanying shape deformations. To this end,
we have established a computational framework that allows us to study the
dynamics of cells using any of the phase-field models that have been
recently developed recently\cite{Shao2012,Ziebert2016,Moure2016,Najem2016}. This
phenomenological approach allows us to easily model the
complex coupling between the shape and motility of the cells, as well
as their interactions with the substrate. In this work, we have adopted a
generic model for fast crawling cells, which is nevertheless able to
reproduce a wide-variety of motility modes seen
experimentally\cite{Lober2014}. Following previous work, which
found a frequency dependent instability in the focal
adhesions\cite{Kong2008,Zhong2011}, and the fact that no orientational
order was observed in the actin network of \textit{Dictyostelium}
under stretching, we have focused our study on the role of the adhesion dynamics on the reorientation. Given the strong frequency dependence found by Kong et
al.\cite{Kong2008}, and the reports of a lower frequency threshold to
observe reorientation (albeit in slow crawling
cells)\cite{Jungbauer2008,Liu2008}, we  assumed a sigmoidal response
for the adhesion dynamics on the rate of
deformation, such that they detach ($d\ne 0$) if the rate at which they are being
deformed exceeds a given threshold. Furthermore, we can selectively
tune this response so that the cells become sensitive only to
compression ($d^{(-)}$) or extension ($d^{(+)}$), or both
($d^{(\pm)}$). Even using this simple coupling we are still able
to obtain a non-trivial frequency dependent reorientation for our model
cells. Depending on whether the cells tend to detach and stop crawling under
too large extension or compression, or whether they are just being
passively advected by the substrate, and how the stretching frequency compares to the
characteristic frequencies associated to the shape deformation and the
actin and adhesion dynamics, we can observe both perpendicular or parallel alignment,
as well as oblique orientations.

As a reference, we considered first the
passive case ($d=0$). At very low frequencies, there is
no reorientation, with the cell oscillating around its initial
direction. As the frequency is increased, both the parallel $\theta=0$ and
perpendicular $\theta=\pi/2$ directions become steady state solutions,
but there is not systematic reorientation (i.e., cells do not show any
preference between either direction). This (slow) reorientation arises
because the actin network can no longer follow the deformations of the
substrate. At higher frequencies, past the characteristic frequency
associated to the shape deformations (as given by the membrane
stiffness), the reorientation time scale is considerably reduced, but
there is still no preference between parallel or perpendicular
directions. Finally, upon a further increase in the stretching
frequency, we reach the time-scales over which the adhesion
attachments are formed. It is at this point where we observe complete
reorientation in the parallel direction. This parallel alignment has
been predicted to occur in cases where the cellular activity is
negligible compared to the forces coming from the
substrate\cite{De2007,De2008}, which is in line with our numerical
predictions. 

In the case of an active coupling with the substrate ($d\ne 0$), 
we observed complete realignment, either in the parallel
or perpendicular directions, over most of the parameter range
considered. Thus, our results provide further evidence for the fact that
the stability of adhesion bonds can have a dramatic effect on the
mechanosensitivity of crawling
cells\cite{Kong2008,Zhong2011,Chen2015a}. For all three
types of responses ($d^{(\pm)}, d^{(+)}, d^{(-)}$), we were able to
observe complete perpendicular alignment, as has been reported
experimentally for fast-crawling
cells\cite{Iwadate2009,Iwadate2013,Okimura2016a}, for low to moderate
frequencies. This is particularly noticeable for the $d^{(+)}$ cells,
for which the perpendicular direction is the stable
orientation over a wide range of frequencies. In contrast, $d^{(-)}$
cells show a preference to align in the parallel direction. Thus, cells that resist
extension (compression) will usually align perpendicular (parallel) to the direction of
stretching. However, at high enough frequencies the
cells always align parallel to the direction of stretching, just as
the passively advected cells ($d=0$).

Our theory predicts that in the case where the adhesion dynamics
dominates the response of the cell, any asymmetry during the
loading/unloading phases of the stretching can be used to align the
cells along directions parallel or perpendicular to the
stretching. This asymmetry can be intrinsic to the cell, if it is able
to respond differently to extension and compression, or it can be due
to the stretching protocol itself. This is relevant with regards to
the experiments reported by Iwadate et
al.\cite{Iwadate2009,Iwadate2013,Okimura2016a}, since they have not
used a sinusoidal signal, with symmetric loading and unloading, but
a saw-tooth like signal, with fast extension followed by a slow
compression. Even if the cell cannot distinguish between
extension and compression, but only the magnitude of the rate of
deformation, this would correspond to our cells with the $d^{(+)}$
response. Indeed, we have shown that for moderate frequencies, these
cells prefer a perpendicular orientation, as reported
experimentally. This could be easily tested by repeating the
experiments with a complementary experimental protocol that had slow extension
followed by fast relaxation. Such a case would correspond to our
$d^{(-)}$ cells, and our theory predicts that the preferred
orientation could then be switched to the parallel direction.  

 Our approach will prove useful to
study the mechanosensitivity of fast crawling cells, since it can
incorporate the salient features: (1) the elastic response of the cell, (2) the
forces on the focal adhesions, (3) the active forces generated by
the cell, and (4) the complex coupling between cell shape and
motility. In addition, the cell-level description we propose can be
trivially extended to multi-cellular systems to study the
mechanosensitivity of tissues. Finally, we would like to point out
that the generic model used here has not been parametrized to any
particular cell type. Thus, more work is required to obtain precise
quantitative comparisons with experiments. This will be the focus of
future investigations, where we will consider a more detailed coupling
between the cell and the substrate, as well as the effect of membrane
tension and substrate elasticity, and how they affect the
actin (de)polymerization rates\cite{Winkler2016}.
\acknowledgments{
  JJM would like to acknowledge fruitful discussions with Natsuhiko
  Yoshinaga, Koichiro Sadakane, Kenichi Yoshikawa, Matthew Turner, Takashi
  Taniguchi, and Simon Schnyder during the preparation of this manuscript.
  This work was supported by the Japan Society for the
  Promotion of Science (JSPS) Wakate~B (17K17825) and KAKENHI
  (17H01083) grants, as well as the JSPS bilateral joint
  research projects.
}
\appendix
\section{Intrinsic Time Derivatives and Conservation Laws}
\label{s:app_tensor}
To modify the equations of motion of the crawling cell for the case
where the substrate itself is being stretched, we need to carefully
translate the formulas to a time-dependent (non-orthonormal)
coordinate system. For the spatial gradient operators, we simply replace
partial derivatives ($\p_{x^i}$) with covariant derivatives
($\Grad_i$), however, the main
issue here is how to handle the time derivatives. The material
derivative should not be used, as it does not yield proper tensorial
quantities. Instead, the intrinsic time-derivative should be employed\cite{Aris1989,Venturi2009}.
It defines tensorial quantities that provide the appropriate
time-variation of arbitrary grade tensors along particle paths in
time-dependent curvilinear coordinates. For scalars ($a$) and vectors ($b^i$),
this intrinsic time-derivative takes the following form
\begin{align}
  \Dt{a} &= \p_t a + \parg{u^k - U^k} \nabla_k a \label{e:dt_a}\\
  \Dt{b^i} &= \p_t b^i + \parg{u^k - U^k} \nabla_k b^i + b^k\nabla_k U^i\label{e:dt_vec}
\end{align}
where $u^k$ refers to the $k$-component of the ``particle'' velocity,
$U^k$ to that of the coordinate-flow (i.e., the coordinate-flow
velocity of the moving grid), and $\nabla_k a = \p_k a$ and $\nabla_k
b^i = \p_k b^i + \Gamma^i_{jk} b^j$ are the components of the
covariant derivatives ($\Gamma^{i}_{jk}$ the connection
coefficients). In this work, given the nature of the deformation we
are interested in, all connection coefficients are zero. However, as the body
basis vectors are not orthonormal, since their length is changing in
time, we do need to differentiate between vectors and 1-forms,
or contravariant and covariant components.

Thus, we see that the advection terms
are proportional to the relative velocity $(\bm{u} - \bm{U})$. In
addition, if $\bm{U}=0$, which corresponds to time-independent
coordinates, we recover the standard material derivative $\textrm{D}_t
\bm{a} = \p_t\bm{a} + \bm{u}\cdot\bm{\nabla}\bm{a}$. In this
work, we consider the special case $\bm{u} = \bm{U}$, for which the
advection term is exactly zero. This corresponds to an idealized
situation of a deformable, yet inelastic substrate. That is, we
impose the large-scale deformation of the substrate and ignore any
deviations caused by the traction forces exerted by the cell (as these
are assumed to be much smaller).

The intrinsic time-derivatives allow us to compute the change in
tensorial quantities along particles paths in time-dependent
curvilinear coordinates. However, when formulating conservation laws,
we must consider the time-variation of extensive (integrated) material
quantities. This is given by the Reynolds transport theorem. Consider the total
amount of $a$ carried by a given material element, which may be
deforming in time. The total change in $a$ is defined as\cite{Venturi2009}
\begin{align}
  \frac{\dd}{\df{t}}\int_{V(t)} a \sqrt{g}\df{\xi}^n &= \int_{V(t)}
  \parg[\bigg]{\Dt{a} + a\nabla_k u^k}\sqrt{g} \df{\xi^n} \label{e:reynolds}
\end{align}
where $V(t)$ is the (time-dependent) domain of the material element
under consideration, $\bm{u}$ its velocity, and $g =
\abs{\determ{g_{ij}}}$ is the determinant of the metric tensor.
\section{Numerical Implementation}
\label{s:app_num}
We outline the numerical method used to solve Eqs.~(\ref{e:rho}-\ref{e:A}).
The differential equations are all of the form
\begin{align}
  \p_t u &= \mathcal{L}(t) u + \mathcal{G}(u,t)
\end{align}
where $\mathcal{L}$ is a linear operator, which can depend on time,
but is independent of $u$, while $\mathcal{G}$ is the non-linear
term. Applying an Euler scheme in time, treating the linear part
implicitly, and the non-linear part explicitly, we have
\begin{align}
  \frac{u_{n+1} - u_n}{h} &= \eta L_{n+1} u_{n+1} + (1-\eta) L_n u_n +
  G_n \\
  u_{n+1} &= \parg[\Big]{1 - h \eta L_{n+1}}^{-1}\parg[\Big]{u_n +
    h\barg[\big]{\parg{1-\eta}L_n u_n + G_n}}
\end{align}
where $u_n = u(t_n)$, $L_n = \mathcal{L}(t_n)$, and $G_n =
\mathcal{G}(u_n, t_n)$, with $h$ the time step and $t_n = n h$. Choosing $\eta = 0$, corresponding to an explicit
calculation of the linear operator, yields
\begin{align}
  u_{n+1} &= u_n + h\barg[\big]{L_n u_n + G_n}
\end{align}
whereas $\eta=1$, corresponding to an implicit treatment, results in
\begin{align}
  u_{n+1} &= \parg[\Big]{1 - hL_{n+1}}^{-1}\parg[\Big]{u_n +
    h G_n}
\end{align}
We use the latter due to its improved stability. To resolve the
differential operators, we employ a pseudo-spectral method\cite{Canuto2006,BulentBiner2017}, solving
the equation of motion in Fourier space, but computing all non-linear
terms in real space and then transforming to Fourier space. For the
equations we have considered, the linear operator is usually just the
diffusion term $L\propto \Delta u$, which in Fourier space is just
$\hat{L}\propto \Norm{k}^2 \hat{u} = G^{IJ}k_I k_J \hat{u}$ (with $k$ the
wave-vector). Fourier transforms were performed using the Fast Fourier
Transform, with a typical grid size of $256\times 256$ points on a square domain
of size $L=100$.
\bibliography{kerato}
\end{document}